\documentclass[aps,prl,superscriptaddress,twocolumn,showpacs,floatfix]{revtex4}
\usepackage{graphicx}
\usepackage{epsf}
\usepackage{bm}
\usepackage{latexsym}
\usepackage{makeidx}
\RequirePackage{amsopn} \RequirePackage{amsfonts}
\RequirePackage{amsthm}

\begin{document}

\title{Excitons in Molecular Aggregates with L\'evy Disorder:
\\ Anomalous Localization and Exchange Broadening of Optical Spectra}

\author{A. Eisfeld}
\affiliation{Max Planck Institute for Physics of Complex Systems,
N\"othnitzer Strasse 38, D-01187 Dresden, Germany}

\author{S. M. Vlaming}
\affiliation{Centre for Theoretical Physics and Zernike Institute
for Advanced Materials, University of Groningen, Nijenborgh 4, 9747 AG
Groningen, The Netherlands}

\author{V. A. Malyshev}
\affiliation{Centre for Theoretical Physics and Zernike Institute
for Advanced Materials, University of Groningen, Nijenborgh 4, 9747 AG
Groningen, The Netherlands}

\author{J.\ Knoester}
\affiliation{Centre for Theoretical Physics and Zernike Institute
for Advanced Materials, University of Groningen, Nijenborgh 4, 9747 AG
Groningen, The Netherlands}


\begin{abstract}

We predict the existence of exchange broadening of optical
lineshapes in disordered molecular aggregates and a nonuniversal
disorder scaling of the localization characteristics of the
collective electronic excitations (excitons). These phenomena occur
for heavy-tailed L\'evy disorder distributions with divergent second
moments - distributions that play a role in many branches of
physics. Our results sharply contrast with aggregate models commonly
analyzed, where the second moment is finite. They bear a relevance
for other types of collective excitations as well.

\end{abstract}


\maketitle

During the past decade much work has been devoted to low-dimensional
optical materials, such as molecular aggregates, photosynthetic
antenna systems, conjugated polymers, and quantum wells and wires.
They derive their interest from strong and color-tunable absorption
and luminescence properties, fast electromagnetic energy transport,
or optical switching~\cite{Scholes06}. Common to this wide variety
of materials is that their properties are rooted in excitonic
eigenstates, collective excitations that consist of a linear
combination of local excited states. The low-dimensional structure,
however, makes these excitations strongly dependent on the presence
of disorder in the site energies and (or) the intersite
interactions. Disorder localizes the excitons and thereby affects
basic properties, such as the optical oscillator strength per state,
the absorption linewidth, and the excitation diffusion constant. The
standard disorder model considers Gaussian or box-like distributions
for the energies or the interactions, which both have a finite
second moment.

In this Letter, we show that disorder models with divergent second
moments give rise to drastically different results for the exciton
states and the collective optical response than the standard ones.
To this end, we study molecular aggregates with site disorder taken
from distributions that have caught interest in a wide range of
physical problems, namely heavy-tailed L\'evy
distributions~\cite{Levy24}. These were introduced originally as
"exceptional cases" that do not obey the central limit theorem.
However, during the past 20 years it has been recognized that they
frequently occur in physics and their divergent second moments give
rise to a variety of surprising effects in subfields ranging from
statistical physics to optics, plasma physics, and condensed matter
physics~\cite{Metzler00,Shlesinger93, Ott90,
Sokolov97,Pereira04,Barthelemy08,Chechkin02,Muller01,Kharlamov02,Barkai00,Barkai03}.
In the context of the model studied in this Letter, it is of
particular interest that molecules embedded in a structurally
disordered (glassy) host provide an interesting realization of
systems with heavy-tailed disorder. The interactions between the
molecule and the multipoles representing the structural disorder of
the host give rise to random shifts of the energy levels of the
molecule, which is reflected in the broadening of the absorption
line. Indeed, it has been shown experimentally and theoretically
that these absorption lines have heavy tails (decaying slower than
Lorentzian)~\cite{Muller01,Kharlamov02}, with mean and variances
obeying L\'evy distributions~\cite{Barkai00,Barkai03}.

In previous studies, L\'evy statistics applied to a single random
variable, such as the displacement of a diffusing particle or
excitation energy of a single molecule. Here we present the first
study of interacting degrees of freedom that individually obey
heavy-tailed L\'evy statistics and predict totally new phenomena.
More specifically, we show that the model considered has remarkable
collective optical properties that differ even qualitatively from
the well-studied case of Gaussian disorder~\cite{Fidder91}. In
particular, we find the phenomenon of exchange broadening of the
absorption line shape, the counterintuitive appearance of fine
structure in the density of states (DOS) and the absorption spectrum
with increasing disorder strength, and a nonuniversal disorder
scaling of the distribution of exciton localization lengths.

We consider a one-dimensional Frenkel exciton model to describe the
optical properties of linear aggregates of dye molecules. In fact,
this model is very generic and also is used often to describe the
properties of the other optical materials mentioned in the
introduction. The model consists of a linear array of two-level
molecules with parallel transition dipoles, which interact through
dipole-dipole transfer interactions. The optical excitations of this
system are described by the eigenstates of the exciton Hamiltonian
matrix $H_{nm} = E_n \delta_{nm} - J (\delta_{m,n+1} +
\delta_{m,n-1})$ \cite{Fidder91}, where for simplicity only
nearest-neighbor couplings are considered~\cite{endnote}. The
transfer integral $J>0$ is assumed to be constant. Disorder is
included by taking the site energies $E_n$ as uncorrelated
stochastic variables, drawn from a distribution $p(E)$.

Special for our model, as compared to previous studies, is that we
consider heavy-tailed L\'evy distributions for $p(E)$. In
particular, we consider the class of symmetric L\'evy distributions
with mean zero, given by \cite{Feller66}
\begin{equation}
\label{Levy distribution}
    p(E) = \frac{1}{2\pi} \int_{-\infty}^{\infty} dt e^{iEt}
    \exp \left( -|\sigma t|^{\alpha} \right) \ .
\end{equation}
Here, $\sigma > 0$ and $0< \alpha \le 2$ are called the scale
parameter and index of stability, respectively. The former
determines the half-width at half maximum (HWHM) of $p(E)$ and we
will hereafter refer to it as the disorder strength, while the
latter fixes the asymptotic behavior $p(E) \sim 1/E^{1+\alpha}$ for
$E \gg \sigma$ and $\alpha < 2$. The power-law behavior for large
$E$ gives rise to heavy tails in $p(E)$, which in turn lead to a
divergent second moment.

It should be noted that the average site energy for a disorder
realization on the chain, ${\bar E} = N^{-1} \sum_{n=1}^N E_n$, also
obeys a L\'evy distribution, with the same index of stability, but a
renormalized disorder strength \cite{Feller66}:
\begin{equation}
 \label{sigma*}
    \sigma^* = \sigma N^{\frac{1-\alpha}{\alpha}} \ .
    \label{scale}
\end{equation}
This property is referred to as stability of the distribution. For a
Gaussian distribution ($\alpha = 2$), Eq.~(\ref{scale}) gives the
well-known result $\sigma^* = \sigma/\sqrt{N}$, which reflects the
exchange narrowing effect of the optical line shape of molecular J
aggregates \cite{Knapp84}: the energy distribution of delocalized
excitons is narrower than $\sigma$, because they average over the
energy fluctuations of the individual sites. By analogy one expects
from Eq.~(\ref{scale}) that for $\alpha \le 1$, this effect is
absent. The Lorentzian case ($\alpha = 1$) is special, as it yields
$\sigma^* = \sigma$. This case was studied in
Refs.~\cite{Eisfeld06}; while it does not exhibit exchange
narrowing, its main properties still are similar to those for
Gaussian disorder. Here, we are particularly interested in
distributions with $\alpha < 1$, for which $\sigma^* > \sigma$
(exchange broadening).

\begin{figure}[ht]
\begin{center}
\includegraphics[width=\columnwidth]{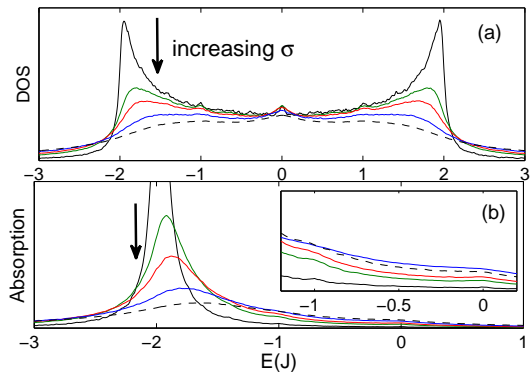}
\end{center}
    \caption{DOS (a) and absorption spectra (b) for
    L\'evy disorder with $\alpha = 1/2$ and disorder
    strengths $\sigma=$ 0.01$J$, 0.1$J$, 0.2$J$, 0.5$J$, and $J$.
    The arrows mark the curves in order of increasing $\sigma$.
    }
\label{fig:DOSandABSandOSS}
\end{figure}

\begin{figure}[ht]
\begin{center}
\includegraphics[width=\columnwidth]{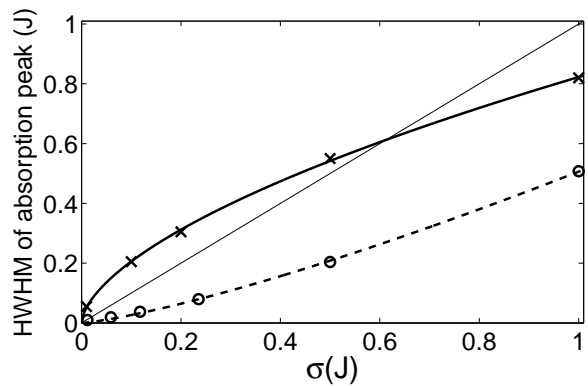}
\end{center}
    \caption{The HWHM of the absorption spectrum as a function of
    $\sigma$ for L\'evy disorder with $\alpha = 1/2$ (crosses) and
    for Gaussian disorder (circles). The corresponding power law
    fits are plotted as a solid and a dashed line, respectively.
    The thin solid line ${\mathrm {HWHM}}=\sigma$ is plotted for
    reference.
    }
\label{fig:HWHMofABS}
\end{figure}

\begin{figure}[ht]
\centerline{\includegraphics[width=0.75\columnwidth]{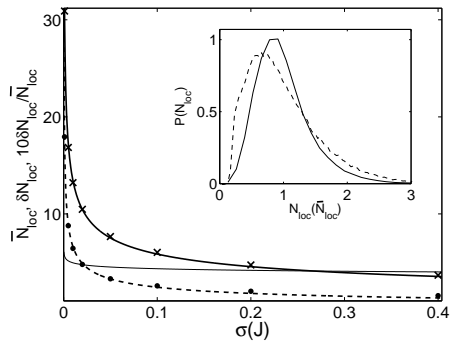}}
    \caption{The average,
    ${\bar N}_\mathrm{loc}$, (crosses) and the standard deviation,
    $\delta N_\mathrm{loc}$, (dots) of the localization length
    distribution $P(N_\mathrm{loc})$ as functions of $\sigma$ for
    L\'evy disorder with
    $\alpha=1/2$. The thick solid and dashed lines are power-law
    fits, respectively (see text). The thin solid line gives
    $10(\delta N_\mathrm{loc}/\bar{N}_{loc})$, while the inset shows
    the localization length
    distributions for $\sigma=0.001J$ (dashed) and for $\sigma=0.1J$
    (solid).
     }
\label{fig:scaling3}
\end{figure}

\begin{figure}[ht]
\begin{center}
\includegraphics[width=\columnwidth]{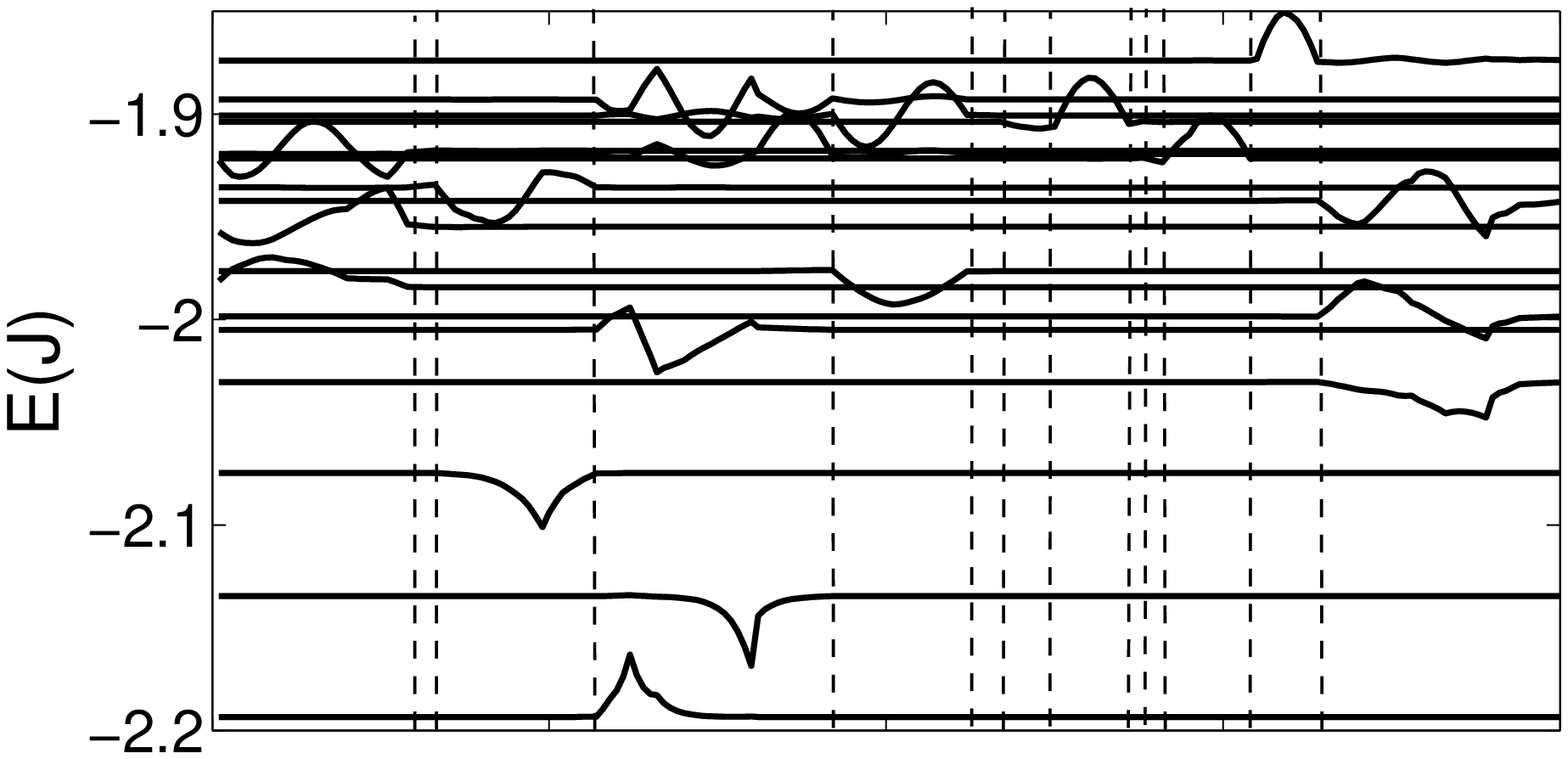}
\vspace{-0.5cm}

\includegraphics[width=\columnwidth]{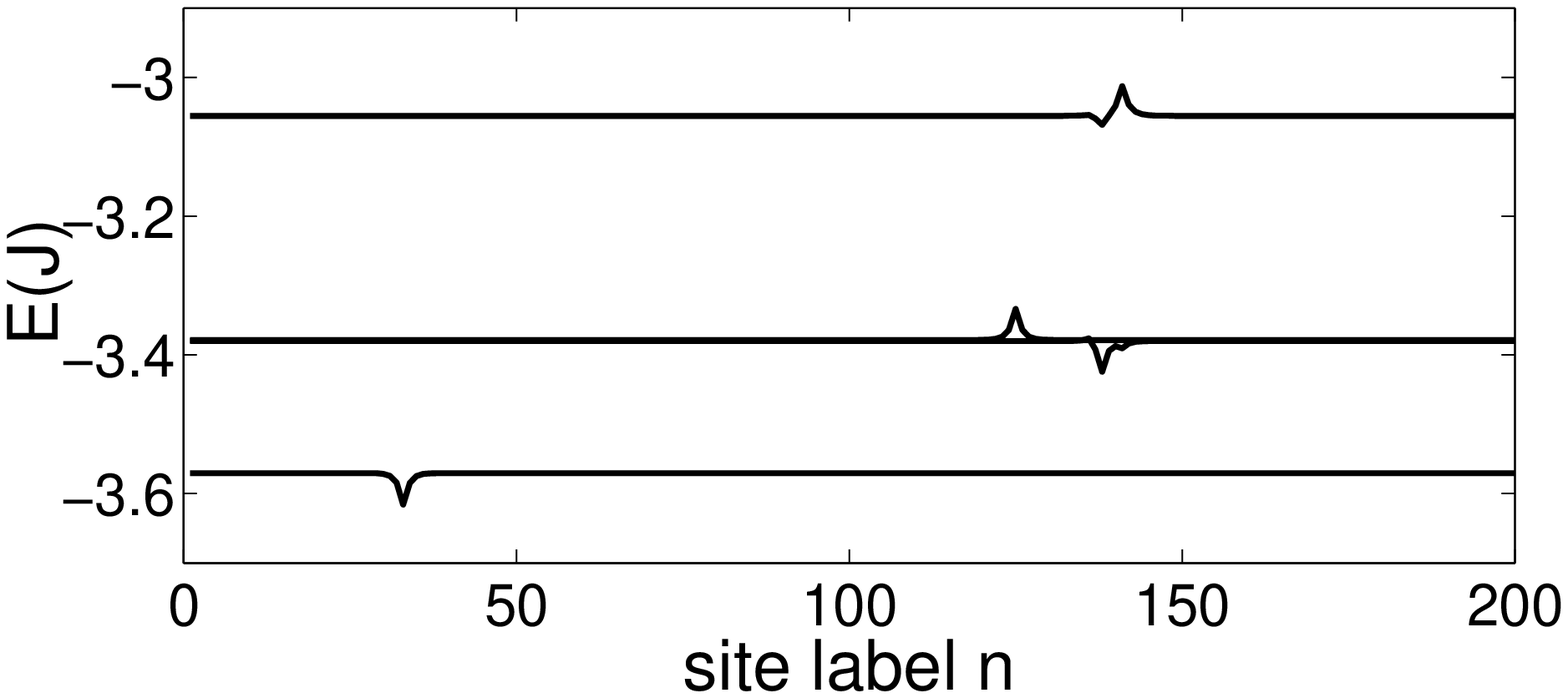}
\vspace{-1cm}
\end{center}
    \caption{Exciton wave functions and energies
    for a typical realization of L\'evy disorder with $\sigma = 0.1 J$ and
    $\alpha=1/2$. The upper panel focuses on energies around the lower band edge;
    the lower panel displays an energy interval deep in the DOS tail.
    The dashed vertical lines indicate the positions of outliers in the site energy.
    }
\label{fig:scaling1}
\end{figure}

We analyzed our model by numerical simulations. We used the
algorithm described in Refs.~\onlinecite{Chambers76} to generate
the random energies ${E_n}$, after which the Hamiltonian was
diagonalized numerically and the exciton DOS, the absorption
spectrum, and the exciton localization lengths $N_\mathrm{loc}$
(inverse participation ratio) were calculated using standard
methods~\cite{Fidder91}. We used chains of 200 sites and averaged
over tens of thousands of disorder realizations. As typical examples,
the results obtained for $\alpha = 1/2$ are presented in
Figs.~\ref{fig:DOSandABSandOSS} -~\ref{fig:scaling1}.

Figure~\ref{fig:DOSandABSandOSS} shows the DOS and the absorption
spectrum for various disorder strengths $\sigma$. The DOS exhibits a
typical one-dimensional (1D) shape, with peaks at the band edges
$E=\pm 2J$ that are smeared by disorder. However, three additional
features are visible: one at the band center, $E=0$, and two more at
$E= \pm J$. The relative importance of these three features strongly
depends on $\sigma$ and $\alpha$, revealing a transition from a 1D
excitonic DOS to a mostly monomeric DOS with increasing $\sigma$.
The origin of the extra features, which we found to get more
pronounced upon decreasing $\alpha$, will be explained below.

The structure in the DOS is also reflected in the absorption
spectrum [Fig.~\ref{fig:DOSandABSandOSS}(b)]. While this spectrum is
dominated by the intense band edge peak (the J band), characteristic
for J aggregates, two much less intense features occur at $E = -J$
and $E=0$. For Gaussian or Lorentzian disorder these features cannot
be discerned. Figure~\ref{fig:DOSandABSandOSS}(b) also reveals
another new effect: with increasing value of $\sigma$, the J band
shifts to the blue, while for Gaussian disorder a red-shift
occurs~\cite{Fidder91}.

Figure~\ref{fig:HWHMofABS} gives the HWHM of the absorption band as
a function of $\sigma$, for $\alpha=1/2$ (crosses) and for
comparison also for $\alpha=2$ (Gaussian; circles). This plot
confirms the most fascinating effect of heavy-tailed L\'evy
distributions, anticipated above, namely the occurrence of exchange
broadening: While for Gaussian disorder the HWHM is smaller than the
bare disorder $\sigma$, for $\alpha=1/2$ it is larger (for $0 <
\sigma < 0.6$). The best power-law fit to the data reads
$\mathrm{HWHM} = 0.85 J \left(\sigma/J\right)^{0.60 \pm 0.03}$
(solid line), which differs strongly from the scaling $\mathrm{HWHM}
\propto J \left(\sigma/J\right)^{4/3}$ for the Gaussian case (dashed
line)~\cite{Fidder91}. While the broadening effect qualitatively
agrees with Eq.~(\ref{scale}), we will see below that the
quantitative explanation of the exponent $0.60 \pm 0.03$ is more
subtle.

Finally, Fig.~\ref{fig:scaling3} characterizes the distribution
$P(N_\mathrm{loc})$ of the localization length of the exciton states
that occur in the energy interval $[-2.1J,-1.9J]$, i.e., around the
lower band edge $E_b=-2J$, the region that dominates the optical
response. Shown are the numerical data (symbols) and their power-law
fits (lines) for the average, $\bar{N}_\mathrm{loc}$, (crosses and
solid line) and the standard deviation, $\delta N_\mathrm{loc}$,
(dots and dashed line) of $P(N_\mathrm{loc})$ as a function of
$\sigma$. The fits represent the data very well, and read
$\bar{N}_\mathrm{loc} = 2.56(J/\sigma)^{0.36}$ and $\delta
N_\mathrm{loc} = 0.91(J/\sigma)^{0.43}$. Thus, the ratio $\delta
N_\mathrm{loc}/ \bar{N}_\mathrm{loc}$ is not constant, implying that, in
contrast to Gaussian and Lorentzian disorder~\cite{Eisfeld06},
L\'evy disorder with $\alpha < 1$ does not lead to a universal
function for $P(N_\mathrm{loc})$ in the $\sigma$-range considered.
This is clearly observed from the inset of Fig.~\ref{fig:scaling3},
where $P(N_\mathrm{loc})$ is plotted for $\sigma=0.001J$ and
$\sigma=0.1J$, after scaling the horizontal axis to the average
localization length appropriate for the $\sigma$ value considered.
We note that the scaling relations described here, only weakly
depend on the choice of the energy interval around $E_b$.

The explanation of all phenomena found above, lies in the interplay
between two mechanisms for localizing the exciton states around
$E_b$ by heavy-tailed L\'evy disorder. For Gaussian disorder only
one such mechanism exists, namely localization of states in
effective potential wells created by random site energies
\cite{Lifshits88}. The typical localization length $N^*$ of these
states can be obtained from a balance of the energy spacing $\sim
J/N^{*2}$ of two adjacent states within a localization segment and
their disorder-induced scattering rate, which scales according to
$\sigma^*$ given in Eq.~(\ref{sigma*}) with $N$ replaced by $N^*$
\cite{Malyshev95}. Generalizing to the case of L\'evy distributions
yields
\begin{equation}
\label{N*}
    N^{*}=\left(\frac{3\pi^2J}{\xi_\alpha \sigma}\right)^{\frac{\alpha}{1+\alpha}},
\end{equation}
where $\xi_\alpha$ is a numerical factor of order unity ($\xi_{1/2}
\approx 0.7)$. For $\alpha = 1/2$, this gives $N^* \propto
(J/\sigma)^{1/3}$. Generally, $N^*$ is not identical to $\bar{N}_\mathrm{loc}$
because the latter also includes effects of segmentation (see below).

For Gaussian disorder, $N^{*}$ is the only length scale relevant to
the band edge states and the optical response. In the case of the
heavy-tailed L\'evy distributions, however, a second localization
mechanism - and corresponding length scale - exists. The long tails
lead to a high concentration of outliers, i.e., sites with energy
$|E_n| > 2J$. These fluctuations are so large that the interaction
$J$ cannot overcome them; they therefore break up the chain in
segments of length $N_\mathrm{seg}$, capped by two outliers, which
form the maximum intervals over which excitons may delocalize. For
decreasing values of $\alpha$ this effect gets stronger. It is
straightforward to show that for $\alpha \leq 1$ the segment length
distribution is exponential, with mean
\begin{equation}
 \label{mean Nseg}
    \bar{N}_\mathrm{seg}
     = \frac{\pi}{2\Gamma(\alpha)\sin(\pi\alpha/2)}
     \left(\frac{2J}{\sigma}\right)^{\alpha} \ .
\end{equation}
$\bar{N}_\mathrm{seg}$ is the second length scale in the problem.

The existence of two localization mechanisms is confirmed by
Fig.~\ref{fig:scaling1}, which for $\alpha=1/2$ and $\sigma = 0.1J$
shows a typical realization of the exciton wave functions and
energies in the neighborhood of the lower band edge and well below
it. Indeed, we observe quite a few outliers, with positions
indicated by the vertical dashed lines. Each outlier has a strongly
localized $s$-like, i.e., without nodes, exciton state (lower panel);
not all these states are seen, as several of them have energies
outside the plot's range. With regards to the two localization
mechanisms, three situations may be distinguished, each of which
occurs in Fig.~\ref{fig:scaling1}.

(i) $N^* < N_\mathrm{seg}$. This is the situation common for
Gaussian disorder. Near the lower band edge one then typically finds
doublets of {\em s}- and {\em p}-type states (with no nodes and one
node, respectively) roughly localized on the same
interval~\cite{Malyshev95}. In Fig.~\ref{fig:scaling1} this
occurs between $n = 52$ and $n = 96$.

(ii) $N^* \approx N_\mathrm{seg}$. Here one finds multiplets of
three or more states on a single segment, which resemble the states
of a disorder-free chain of size $N_\mathrm{seg}$. In
Fig.~\ref{fig:scaling1} examples occur between $n = 1$ and $n = 31$,
between $n = 92$ and $n = 112$, and between $n = 164$ and $n = 200$.

(iii) $N^* > N_\mathrm{seg}$. In this case the segmentation strongly
confines the excitons and the states within a segment typically get
further separated than in the absence of segmentation. If the
segments are still relatively large, their lowest {\em s}-like
eigenstates contain considerable oscillator strength and occur just
above the band edge (e.g., the state between $n=154$ and $n=164$ in
Fig.~\ref{fig:scaling1}). With increasing value of $\sigma$, the
segments get shorter and the energy of the {\em s}-like state grows.
This explains the blue-shift of the J band found in
Fig.~\ref{fig:DOSandABSandOSS}(b). Also, for growing $\sigma$,
segments of length 1 and 2 become more likely. These give rise to
states with energies distributed around the average monomer and
dimer energies, $E = 0$ and $E = \pm J$, respectively, thus
explaining the extra features in the DOS and the absorption spectra
observed in Fig.~\ref{fig:DOSandABSandOSS}. Closer scrutiny even
reveals features for segments of length 3, 4, etc, but these are
weak and overshadowed by the band edge peaks. As is seen in
Fig.~\ref{fig:DOSandABSandOSS}, all these extra features are washed
out when $\sigma$ approaches $J$, because they broaden and cannot be
distinguished anymore.

Of course, the existence of two localization mechanisms, each with
its own length scale, is directly responsible for the observed
nonuniversality of the localization length distribution. Moreover,
the interplay between both mechanisms also explains the disorder
scaling of the exchange broadening of the absorption band. Using
Eq.~(\ref{sigma*}) and replacing $N$ by either $N^*$ or ${\bar
N}_\mathrm{seg}$, we obtain the contributions to the HWHM from the
states localized by the two different mechanisms, respectively. This
leads to $\mathrm{HWHM}^* \propto J(\sigma/J)^{2\alpha/(1+\alpha)}$
and $\mathrm{HWHM}_\mathrm{seg} \propto J(\sigma/J)^{\alpha}$. For
$\alpha = 1/2$, the scaling exponent equals 2/3 and 1/2,
respectively. Since the numerically obtained exponent equals $0.60
\pm 0.03$, we conclude that both scales, $N^*$ and
$\bar{N}_\mathrm{seg}$, almost equally contribute to the disorder
scaling of the J band width.

In summary, we have shown that heavy-tailed L\'evy disorder has
dramatic consequences for the optical properties of linear molecular
aggregates. Novel effects occur, such as exchange broadening and a
nonuniversal scaling of the distribution of exciton localization
lengths. We have shown that these effects may all be traced back to
the simultaneous occurrence of two different localization
mechanisms, each with its own length scale.  We expect that the
occurrence of two localization length scales has equally dramatic
effects on other collective excitations subjected to heavy-tailed
disorder.


\begin{thebibliography}{99}

\bibitem{Scholes06} For a recent overview, see G.D. Scholes and G. Rumbles,
    Nat. Materials {\bf 5}, 683 (2006).

\bibitem{Levy24} P. L\'evy, Bull. Soc. Math. France {\bf 52}, 49 (1924).

\bibitem{Metzler00} R. Metzler and J. Klafter, Phys. Rep. {\bf 339}, 1
    (2000).

\bibitem{Shlesinger93} M. F. Shlesinger, G. M. Zaslawski, and J.
    Klafter, Nature {\bf 363}, 31 (1993).

\bibitem{Ott90} A. Ott {\it et al.},
    Phys. Rev. Lett. {\bf 65}, 2201 (1990).

\bibitem{Sokolov97} I. M. Sokolov, J. Mai, and A. Blumen, Phys. Rev.
    Lett. {\bf 79}, 857 (1997).

\bibitem{Pereira04} E. Pereira {\it et al.}, Phys. Rev. Lett. {\bf 93},
    120201 (2004).

\bibitem{Barthelemy08} P. Barthelemy, J. Bertolotti, and D. S. Wiersma,
    Nature {\bf 453}, 495 (2008).

\bibitem{Chechkin02} A. V. Chechkin, V. Yu. Gonchar, and M. Szydlowski,
    Phys. of Plasmas {\bf 9}, 78 (2002).

\bibitem{Muller01} J. M\"uller, D. Haarer, and B. M. Kharlamov,
    Phys. Lett. A {\bf 281}, 64 (2001).

\bibitem{Kharlamov02}  B. M. Kharlamov and G. Zumofen, J. Chem. Phys.
    {\bf 116}, 5107 (2002).

\bibitem{Barkai00} E. Barkai, R. Silbey, and G. Zumofen, Phys. Rev.
    Lett. {\bf 84}, 5339 (2000).

\bibitem{Barkai03} E. Barkai {\it et al}, Phys. Rev. Lett. {\bf 91},
    075502 (2003).

\bibitem{Fidder91} M. Schreiber and Y. Toyozawa, J. Phys. Soc.
    Jpn. {\bf 51}, 1528 (1982); H. Fidder, J. Knoester,
    and D. A. Wiersma, J. Chem. Phys. {\bf 95}, 7880 (1991).

\bibitem{endnote} We have checked numerically that accounting for the
    long-range dipolar coupling between molecules does not quantitatively
    change the effects found by us, but slightly changes the scaling exponents
    and prefactors, as well as the values for $\alpha$ and $\sigma$ where
    nonuniversality sets in.

\bibitem{Feller66} W. Feller, {\it An Introduction to Probability
    Theory and its Applications}, Wiley, New York (1966).

\bibitem{Knapp84} E. W. Knapp, Chem. Phys. {\bf 85}, 73 (1984).

\bibitem{Eisfeld06} (a) A. Eisfeld and J. S. Briggs, Phys. Rev. Lett.
    {\bf 96}, 113003 (2006); (b) S. M. Vlaming, V. A. Malyshev, and
    J. Knoester, Phys. Rev. B {\bf 79}, 205121 (2009).

\bibitem{Chambers76} J. M. Chambers, C. L. Mallows and B. W.
    Stuck, J. Am. Stat. Ass. {\bf 71}, 340 (1976); J. H. McCulloch,
    Bull. London Math. Soc. {\bf 28}, 651 (1996).

\bibitem{Lifshits88} I. M. Lifshits, S. A. Gredeskul, and
     L. A. Pastur, {\it Introduction to the Theory of Disordered
     Systems}, Wiley: New York, 1988.

\bibitem{Malyshev95} F. Dom\'{i}nguez-Adame and V. A. Malyshev, Am. J.
    Phys. {\bf 72} 226 (2004).





\end{thebibliography}
\end{document}